\documentclass[sigconf,nonacm]{acmart}

\AtBeginDocument{%
  \providecommand\BibTeX{{%
    \normalfont B\kern-0.5em{\scshape i\kern-0.25em b}\kern-0.8em\TeX}}}

\acmConference[]{}{}{}
\acmPrice{}
\acmISBN{}

\usepackage{soul}
\usepackage{xcolor}

\begin{document}

\title{Designing an Artificial Immune System inspired Intrusion Detection System}


\author{William Anderson}
\affiliation{%
  \institution{Mississippi State University}
  \country{Mississippi State, MS, USA}
  }
\email{wha41@msstate.edu}

\author{Kaneesha Moore}
\affiliation{%
  \institution{Mississippi State University}
  \country{Mississippi State, MS, USA}
  }
\email{kkm267@msstate.edu}

\author{Jesse Ables}
\affiliation{%
  \institution{Mississippi State University}
  \country{Mississippi State, MS, USA}
  }
\email{jha92@msstate.edu}

\author{Sudip Mittal}
\affiliation{%
  \institution{Mississippi State University}
  \country{Mississippi State, MS, USA}
  }
\email{mittal@cse.msstate.edu}

\author{Shahram Rahimi}
\affiliation{%
  \institution{Mississippi State University}
  \country{Mississippi State, MS, USA}
  }
\email{rahimi@cse.msstate.edu}

\author{Ioana Banicescu}
\affiliation{%
  \institution{Mississippi State University}
  \country{Mississippi State, MS, USA}
  }
\email{ioana@cse.msstate.edu}

\author{Maria Seale}
\affiliation{%
  \institution{U.S. Army Engineer Research and Development Center (ERDC)}
  \country{Vicksburg, MS, USA}
  }
\email{maria.a.seale@erdc.dren.mil}

\renewcommand{\shortauthors}{Anderson et al.}

\begin{abstract}

The Human Immune System (HIS) works to protect a body from infection, illness, and disease. This system
can inspire cybersecurity professionals to design an Artificial Immune System (AIS) based Intrusion Detection System (IDS). These biologically inspired algorithms using Self/Nonself and Danger Theory can directly augment IDS designs and implementations. In this paper, we include an examination into the elements of design necessary for building an AIS-IDS framework and present an architecture to create such systems. 
  
\end{abstract}



\maketitle

\vspace{-4mm}
\section{Introduction \& Background}

The Human Immune System (HIS) is a large network of organs, white blood cells, proteins (antibodies) and chemicals. This system works together to protect a body from bacteria, viruses, parasites, and fungi that can cause infection, illness, and disease. It is able to successfully monitor, isolate, and mitigate threats 
which provides the system with its version of `threat intelligence' about zero-day `vulnerabilities'. 
This complex system can inspire cybersecurity professionals to design better security systems. Intrusion Detection Systems (IDS) would greatly benefit from immune-system-inspired modifications. 

With the exponential rise of Artificial Intelligence (AI) solutions, `bio-inspired' AI models allow professionals to close the gap between their current security systems and the systems they need to implement.
Much like the human body, an Artificial Immune System (AIS) adapts to growing threats and produces more accurate detectors as necessary. One can draw parallels between the specific advantages of AIS and the needs of cybersecurity professionals designing an IDS. 
{AIS design algorithms using the Self/Nonself model and Danger Theory can directly augment IDS designs and implementations.} 



AIS are a biologically inspired design paradigm based on the Human Immune System (HIS). Broadly, the HIS can be broken down into two distinct layers: \textit{Innate Immune System} and \textit{Adaptive Immune System} \cite{informedhealth.org_2020}. The Innate Immune System is a broad, first response to a foreign substance in the body. This includes physical layers such as skin and mucous membranes, as well as invoked responses, such as inflammation and fever. As the name implies, these are innate to all humans and are a type of nonspecific immune response. The Adaptive Immune System is a second-order immune response that is a specific, targeted response as a means of isolating the pathogen. It is composed of T lymphocytes (T cells), B lymphocytes (B cells), and antibodies. The T cells are responsible for activating other immune responses, the detection of infected cells, and memorization of the pathogen and subsequent immune response. B cells that match the pathogen are recruited by the T cells to multiply and transform into plasma cells. The plasma cells produce antibodies specific to that pathogen; therefore, only the exact antibodies needed are produced. Antibodies have three primary functions: the neutralization of the pathogen (directly or indirectly), to activate other immune system cells (co-operation), and to activate proteins that assist with the immune response. In this way, antibodies support and enhance the response of the innate immune system and adaptive immune system. The exact mechanism for how this process occurs is still under some debate. There are currently two predominate models of the immune system: \textit{Self/Nonself} and \textit{Danger Theory}. 

The Self/Nonself model \cite{lederberg1959genes, burnet1959clonal} espouses that the primary mechanism of this immuno-pathway is determined by the differentiation of self and nonself. During the process of T/B cell creation, the candidate cells are vetted against any proteins contained in the thymus/bone marrow, respectively. Consequently, any that respond to those proteins are destroyed. This self-protection mechanism helps to prevent the immune system from attacking anything that would be deemed self, and in turn, only allows cells that do not respond to self to circulate. This model of immunology has been the primary view since the mid-1950s. However, criticisms against it have led immunologists to seek other explanations, particularly due to a the failure of self/nonself to capture a complete model of the immune system.

\begin{figure*}[ht]
     \centering
     \includegraphics[scale=1.05]{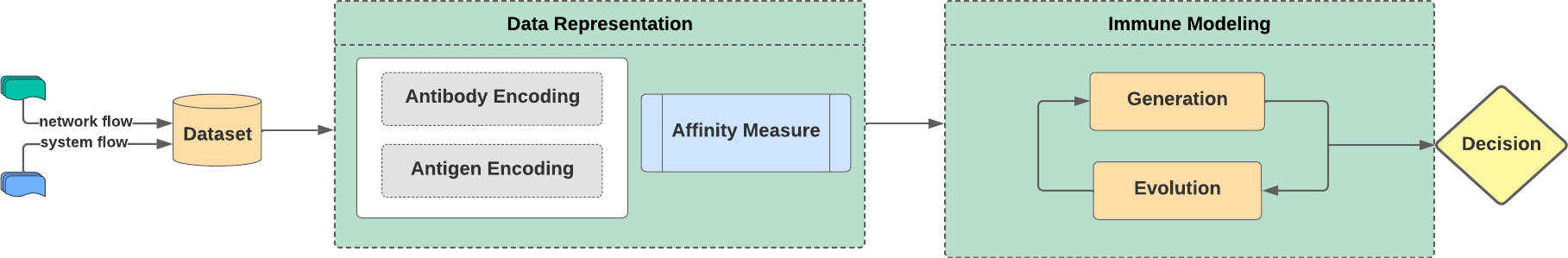}
     \caption{Artificial Immune System inspired Intrusion Detection System (AIS-IDS) Architecture Diagram.}
     \label{fig:arch}
     \vspace{-2mm}
\end{figure*} 

Danger Theory, popularized by Matzinger \cite{Matzinger1994ToleranceDA}, came to prominence in the early-1990s, as an alternative model that arose out of the criticisms of self/nonself. Particularly, criticisms relating to the notion that an immune response is triggered for all nonself entities, whereas self entities would not induce such a reaction. Danger Theory states that the immune system, instead of differentiating self and nonself, differentiates on the basis of danger signals. These signals are released from the body's cells as a means to indicate the presence of danger. For example, in the process cellular recycling, an apoptotic cell death would not create a danger signal, but cellular necrosis would. This addresses the aforementioned criticism by allowing for the immunological model to tolerate nonself entities (e.g. fetus, bacteria) when they present no threat and for an immune response to be elicited from self entities (e.g. cell stress, grafts). Modern research in Danger Theory redefines "danger" to "damage" \cite{Matzinger2002TheDM}, and analyzes flaws on the basis of how it handles innate immunity, cancer, grafts, and exogenous entities \cite{Pradeu2012TheDT}.

Both these models described above can help cyber defense system designers create AIS inspired IDS systems. 
We in the next section, examine the landscape of AIS as it relates to the development of IDS, including both self/nonself and danger theory based approaches.


\section{AIS inspired IDS architecture}

To create an IDS system based on the AIS (henceforth referred to as \textit{AIS-IDS}), we will include an examination into the elements of design necessary for building an AIS-IDS framework: \textit{Data Representation} and \textit{Immune Modeling}, along with potential research recommendations. Figure \ref{fig:arch} includes the overall architecture for a two-staged AIS-IDS framework. The first stage is Data Representation, supported by dataset composition, reflecting how antigens/antibodies are encoded along with the notion of similarity measurement. The second stage is Immune Modeling, which represents the immune theory approach as well as how the AIS-IDS is to evolve over time.

\vspace{-2mm}
\subsection{Data Representation} \label{datarep} 
Ultimately, an IDS concerns itself with the discrimination of normal and anomalous behavior. The first step in that process is the way in which the input data is \textit{represented}. In the traditional AI based IDS, input data is cleaned and prepossessed (e.g. one-hot encoding categorical variables, normalizing continuous variables, and feature selector algorithms) in order to better represent the input data to the problem space. AIS-IDS largely follows this same approach, albeit with a consideration placed on how to best represent the data for both measurements of similarity (affinity) and the subsequent generation algorithm.

\begin{figure}[ht]
     \centering
     \includegraphics[scale=0.6]{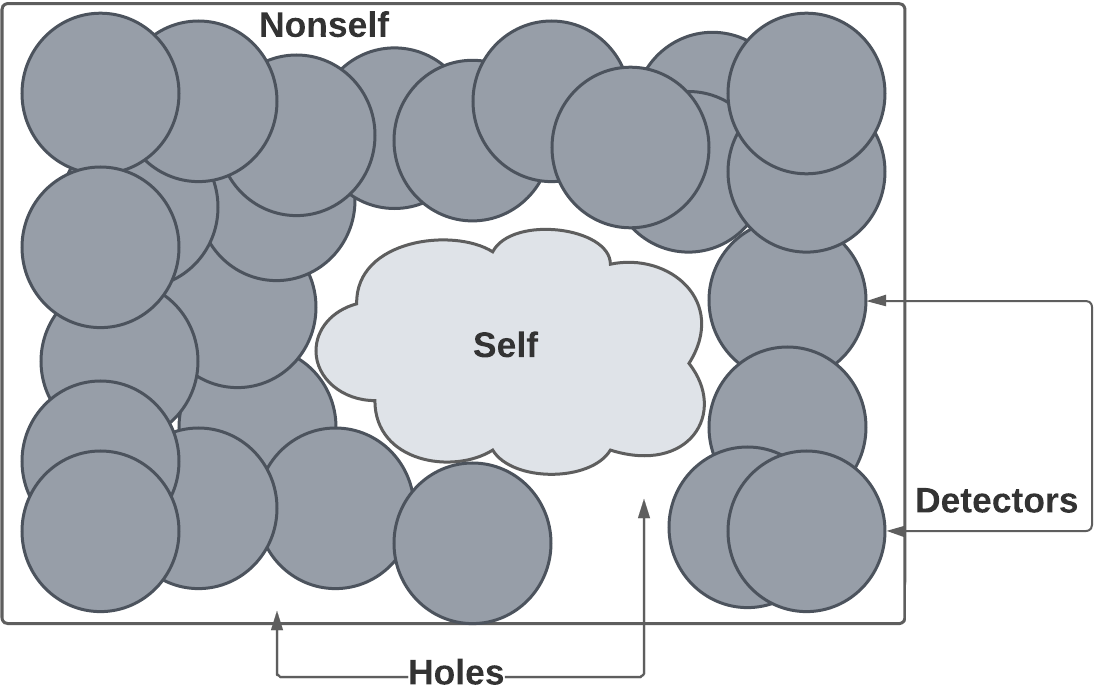}
     \caption{Negative Selection Algorithm - The candidate detectors, generated randomly, are eliminated if they overlap with self. The ideal set of detectors cover the nonself space and leave minimal holes.}
     \label{fig:neg_sel}
     \vspace{-2mm}

\end{figure}

Under the self/nonself approach to AIS-IDS, the core component is the antibody (detector). The antibody detects anomalous data by attempting to match itself to antigens (input data). Therefore, we need to construct a method of representation that allows for the notion of similarity to be calculated. Traditionally, the domain of AIS has accomplished this by representing input data as bit-strings \cite{Forrest1994SelfnonselfDI}. However, this approach has two fundamental issues: affinity measures on bit-strings cover the problem space poorly and there is an exponential growth of computational time with respect to the detector count. More recently, works such as \cite{Gonzlez2002CombiningNS} have approached the antibody/antigen problem not by using bit-strings, but real-valued attributes. By using real-valued attributes, the antibodies can be represented as hyper-shapes, as depicted in Figure \ref{fig:neg_sel}. This allows the representation of the antibody to better cover the non-self space with fewer detectors, solving both the coverage and computational cost flaw of bit-strings.

To create an AIS-IDS using Danger Theory, along with a form of antigen representation, we also need a way to measure `danger'. Danger in the context of immunology are signals such as necrotic cells, pathogenic associated molecular patterns (PAMPs), or inflammatory cytokines. These danger signals, alone, are not enough to be indicative of a systemic problem but would accompany one. However, these danger signals would be formed from the cause-and-effect relationship of the action or presence of the antigen. For IDS, this means that the antigens are best modeled as a `system context' or a snapshot of the state of the to-be classified antigen (e.g. PID, TCP/IP packet fields). Danger signals would then be the measurements of the state of the system over time (e.g. frequency of context switches, number of attempted access violations). Together, these features and historical context are used to establish a matter-of-fact relationship to identify which antigens indicate the source of the danger signals (the intrusion).

\vspace{-3mm}
\subsection{Immune Modeling} \label{immunemodel}
After creating a data representation for an AIS-IDS, we have to consider various choices on modelling the Human Immune System (HIS) for AIS. The notion of AIS-based research began in the mid-1980s with Farmer et al. \cite{Farmer1986TheIS}. Their research proposed that computer science should adopt immune system based modeling which includes the ability to learn, have memory, and be capable of pattern recognition. Their model employed many ideas still in use today, such as a basis of self/nonself, bit-string encoding, and a rudimentary form of negative selection. Currently, an AIS-IDS should consider two modeling choices: \emph{Generation} and \emph{Evolution}. Generation considers the immune theory selection and how detectors and/or danger are formed. Evolution considers how detectors evolve over time, if selected. This enables the AIS-IDS to function as a \textit{dynamic detection system}, robust with respect to polymorphic and zero-day threats, rather than traditional static notions of IDS. 

\vspace{-2mm}
\subsubsection{Generation} \label{generation}
As previously mentioned, the generation model considers which immune theory is being selected and serves as the core for how the detectors are formed. For self/nonself, there are two primary generation algorithms: Negative Selection Algorithm (NSA) \cite{Forrest1994SelfnonselfDI} and Clonal Selection Algorithm (CSA) \cite{Castro2002LearningAO}. NSA, considered to be the seminal algorithm for AIS, is an abstracted model of the biological negative selection process at the core of the self/nonself paradigm. It generates random detectors to identify antigens. These detectors are then evaluated against self, and any which respond are discarded. Thus, the final set contains detectors which only react to non-self antigens, giving rise to very low false positive rates. NSA has many variants such as bit-string NSA, real-valued NSA, v-detector NSA, and hybrid approaches. 

On the other hand, the Clonal Selection Algorithm (CSA), introduced by Castro \& Von Zuben \cite{de2000clonal} and subsequently formalized by them as CLONALG \cite{Castro2002LearningAO}, is an abstracted model of the clonal selection process of the adaptive immune system. CLONALG, at its core, is simply an evolutionary algorithm that mutates detectors on the basis of classification effectiveness. CLONALG has been shown to lower false positive rates \cite{Tang2010AviditymodelBC} and was the basis for the famous aiNet \cite{Castro2000AnEI}. Another algorithm based on the clonal selection process is AIRS \cite{Watkins2004ArtificialIR}, which restricts the evolution process to one-shot, enforcing a resource constraint. Recently, with the rise in popularity of danger theory, algorithms modeled under that framework have met with success, the most popular of which is the Dendritic Cell Algorithm (DCA) \cite{Greensmith2006DendriticCF}. DCA is based on the core principles of danger theory and the antigen process of dendrite cells in humans. Fundamentally, DCA works by fusing multiple signals and performing a correlation on them to determine if-and-when danger is occurring.

\vspace{-1mm}
\subsubsection{Evolution} \label{evolution}
If the development of AIS-IDS is to mimic the HIS, then consideration needs to be given as to how detectors evolve. Given that resources are finite, an infinite number of detectors is implausible. Moreover, the threat landscape is constantly evolving, and so too must the system if it is to be effective in capturing polymorphic and zero-day threats. Therefore, a degree of pruning and generation should be a component of AIS-IDS design to serve as the `living' adaptable component of the system.

The fundamental basis for how this can be accomplished is through the detector. Kim et al. \cite{Kim2002ImmuneMI} extended DynamiCS to remove detectors that were no longer valid or had poor tolerance to new antigens. Pruning detectors that have poor tolerance to a changing landscape is essential to keeping the system robust. However, many times the detectors will need to be replaced. Rather than generating new detectors at random, feedback from previous generations of detectors can help pre-seed new detectors to bootstrap the creation of new detectors. The previous authors extended their system again \cite{Kim2002AMO}, supporting the notion of a gene library, whereby detectors are `remembered' and used to mark past encounters/attacks.

\vspace{-2mm}
\section{Conclusion \& Future Work}

In this paper, we present an architecture for an AIS inspired IDS system. We included an examination of the components necessary for building an AIS-IDS: Data representation and Immune modeling. Our future work includes creating an AIS-IDS system capable of detecting attacks in an organizational setting. 

\vspace{-2mm}
\section*{Acknowledgement}

Work supported by U.S. Department of Defense grant through U.S. Army Engineer Research and Development Center (ERDC) (\#W912HZ-21-C0058) and National Science Foundation (\#1565484, \#2133190). The views and conclusions are those of the authors.

\bibliographystyle{unsrt}
\bibliography{refs}

\end{document}